\documentclass[prl,twocolumn,showpacs,amsmath,amssymb]{revtex4-1}
\usepackage{graphicx}       
\usepackage{mathtools}
\usepackage{epsfig}
\usepackage{dcolumn}        

\newcommand{\beq}{\begin{equation}}
\newcommand{\eeq}{\end{equation}}
\newcommand{\beqa}{\begin{eqnarray}}
\newcommand{\eeqa}{\end{eqnarray}}

\begin{document}
\title{Intrinsic instability of electronic interfaces with strong Rashba coupling}
\author{S. Caprara$^{1,2}$, F. Peronaci$^1$, and M. Grilli$^{1,2}$}
\affiliation{$^1$Dipartimento di Fisica, Universit\`a di 
Roma ``La Sapienza'', P.$^{le}$ Aldo Moro 5, 00185 Roma, Italy}

\affiliation{$^2$ISC-CNR and Consorzio Nazionale Interuniversitario per le Scienze Fisiche della 
Materia, Unit\`a di Roma ``Sapienza''}

\begin{abstract}
{We consider a model for the two-dimensional electron gas formed at the interface 
of oxide heterostructures, which includes a Rashba spin-orbit coupling proportional 
to the electric field perpendicular to the interface. Based on the standard mechanism 
of polarity catastrophe, we assume that  the electric field has a contribution proportional to the 
electron density. Under these simple and general assumptions, we show that a phase 
separation instability (signaled by a negative compressibility) occurs for realistic 
values of the spin-orbit coupling and of the electronic band-structure parameters. This provides an intrinsic 
mechanism for the  inhomogeneous phases observed at the LaAlO$_3$/SrTiO$_3$ or LaTiO$_3$/SrTiO$_3$ interfaces.}
\end{abstract}
\date{\today}
\maketitle

The observation of a two-dimensional (2D) metallic state  at the heterointerface
of two insulating oxides [LaAlO$_3$/SrTiO$_3$ (LAO/STO)] \cite{ohtomo} remarkably 
discovered a new class of high-mobility  electron gases (EGs) important both for fundamental and applicative reasons.
The occurrence of superconductivity in this 2DEG
\cite{reyren,triscone}       and in  LaTiO$_3$/SrTiO$_3$ (LTO/STO) 
\cite{espci1,espci2},  with the possibility to tune the charge density by  gating, 
 has further attracted great attention. On the other hand, there is 
increasing evidence that electron inhomogeneity plays a relevant role in these systems. 
Not only the large width of the superconducting transition in transport experiments is 
a clear indication of charge inhomogeneity \cite{CGBC,BCCG}, but also magnetometry and tunnelling
experiments \cite{ariando,bert,luli,salluzzo2} find submicrometric inhomogeneities. 
While impurities, defects, and other extrinsic mechanisms (see, e.g. Ref. \cite{bristowe})
surely induce inhomogeneities, 
the recent discovery of negative compressibility in a low filling regime \cite{mannhart} 
is a stringent demonstration that an {\it intrinsic} mechanism (like, e.g., an effective 
attraction) is at work to render these 2DEGs inhomogeneous by charge segregation and 
phase separation even in a perfectly clean and homogeneous system. Moreover, even if such 
mechanism were not strong enough to drive the system unstable, it would increase the 
charge susceptibility, emphasizing the effects of the extrinsic mechanisms (impurities, defects 
and so on). 

In this Letter, we point out that a generic source of phase separation is provided by the 
Rashba spin-orbit coupling (RSOC), whenever the electric field determining this coupling 
also controls the electron density. This is precisely the case of LAO/STO and LTO/STO (generically, 
LXO/STO) interfaces, where the two following conditions remarkably meet: i) a 
strong electric field occurs perpendicular to the interface because of the polarity 
catastrophe \cite{polarity,popovic,hirayama} and brings the electrons at the oxide 
interface to produce the 2DEG. An additional electric field, although quantitatively 
less important, is introduced by the gating potential and tunes the density of 
the 2DEG; ii) the parity symmetry is broken at the interface naturally entailing a 
RSOC, which experiments have found to be substantial \cite{cavigliaprl,fete}. We 
show that these two concomitant conditions are enough to drive the 2DEG unstable 
towards phase separation, thereby providing a {\it general and intrinsic} mechanism 
for the inhomogeneity of these oxide interfaces.

The mechanism for RSOC-induced phase separation is rather simple. In a metallic 
system with a rigid band structure the chemical potential $\mu$ increases upon 
increasing the electron density $n$ and the compressibility 
$\kappa\equiv\partial n /\partial \mu$ is positive. On the other hand, if the 
band structure is modified by the charge density (like, e.g., in strongly 
correlated systems, where the quasiparticle bandwidth increases moving away from 
the half-filled Mott-Hubbard insulator) the possibility may occur that $\mu$ 
decreases with increasing $n$ and $\kappa<0$, with the negative compressibility 
region signaling a phase separation. The mechanism at work is so generic 
that it could be relevant for many other systems
in the presence of RSOC like, {\it e.g.}, surface states of topological insulators, 
InAs or GaAs semiconductor heterostructures, or heavy metal surface alloys.

Our analysis finds that
a band structure with large (even anisotropic) masses together with a rather isotropic
Rashba coupling are favorable conditions for this instability to occur.
We also stress that the charge inhomogeneity within the interface is 
due to an inhomogeneous electronic reconstruction and therefore it is
balanced by a correspondingly inhomogeneous redistribution of the countercharges on the top layer.
Therefore the inhomogeneity is not prevented by the standard 
Coulomb mechanisms leading to frustrated phase separation in other systems 
\cite{emerykivelson,FPS,CDG}. This naturally leads to large submicrometric 
inhomogeneities, like those detected in LXO/STO.

\begin{figure}
\includegraphics[angle=0,scale=0.2]{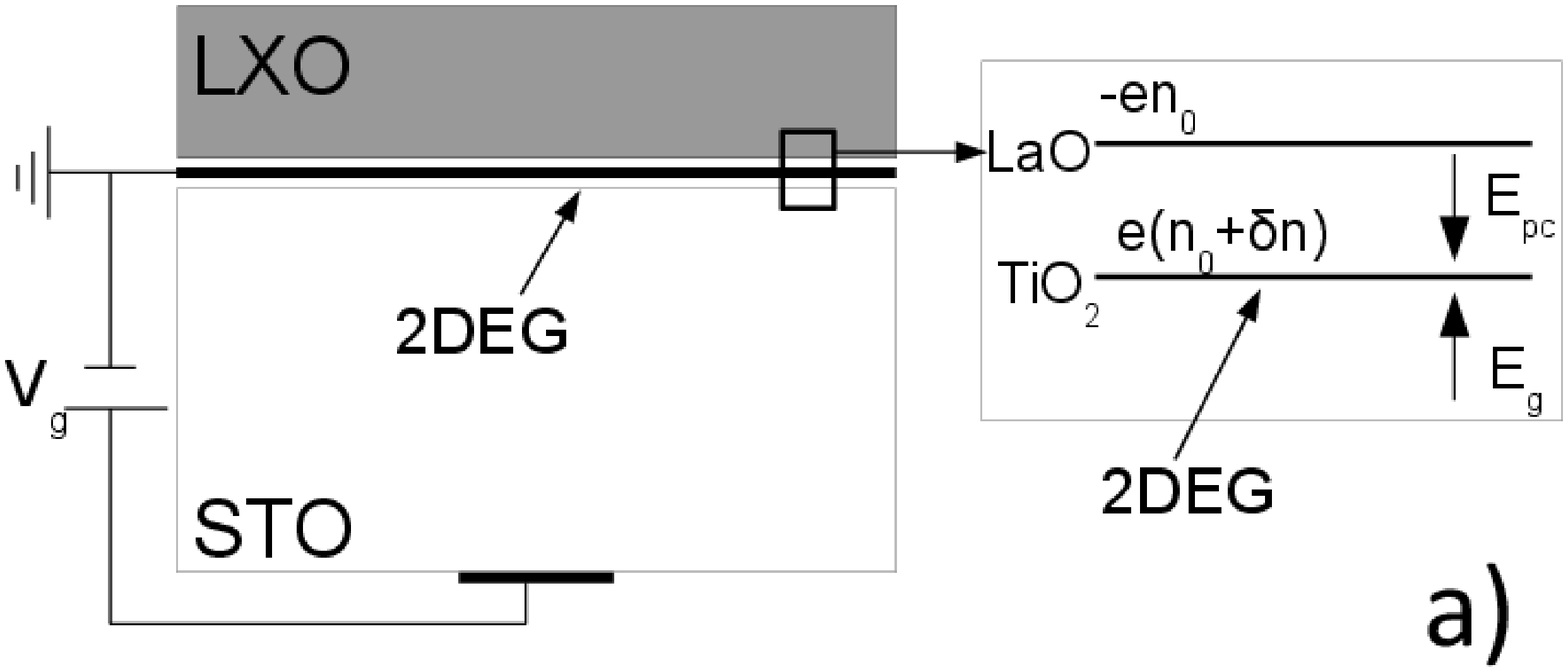}
\includegraphics[angle=0,scale=0.3]{./CPG-fig1bc.eps}
\caption{(Color online) (a) Schematic view of the LXO/STO interface in the presence of gate potential $V_g$. $n_0$ is 
the density of the electrons transferred to the interface by the polarity catastrophe and giving rise the 
$E_{PC}$ electric field. $\delta n$ is the electron density tuned by the gating field $E_g$.
(b) Schematic band structure for the $t_{2g}$ bands along $k_x$ with  $\nu=30$ and $\eta=1$ and (c) $\eta=1/\nu$.
We report the $d_{xy}$ (solid black), $d_{xz}$ (dashed blue online), $d_{yz}$ (dot-dashed, red online) bands.}
%
\label{fig-dos}
\end{figure}

Fig.~\ref{fig-dos}(a) schematically 
shows the 2DEG between the two oxide layers, the gating layers and the resulting 
electric fields. The electron reconstruction to avoid the polarity catastrophe is also depicted with the electron charge 
density $n_0$ transferred onto the interface, leaving oppositely charged planes in 
the LXO layers above it. As a consequence, a field 
$E_{PC}=en_0/(\epsilon_0\epsilon_1)$ arises perpendicular to the interface, where $\epsilon_1\sim 20$ is 
the LXO dielectric constant, $\epsilon_0$ is the vacuum dielectric permittivity and $e$ 
is the electron charge. 
Typical values of $E_{PC}$ may reach $10^8$\,V/m. The total electric field perpendicular 
to the interface is obtained by adding to $E_{PC}$ the gate electric field 
$E_g=V_g/d$, where $d$ is the width of the STO substrate, which tunes the 
charge density with a term $e\delta n=\epsilon_0\epsilon_2E_g$. For a STO substrate 
$0.5 \times 10^{-3}$\,m thick, $E_g\sim 10^5$\,V/m ($\ll E_{PC}$).
It is quite important to notice that the large value of the low-temperature STO dielectric constant 
$\epsilon_2\gtrsim 10^4$ allows for relatively large variations of the electron filling $\delta n$, 
while the gate electric field always stays much smaller than $E_{PC}$. As a consequence, the 2DEG can 
be greatly depleted when $\delta n \sim -n_0$, while the total electric field is still large 
$E=E_g+E_{PC}\approx E_{PC}$.

For a description of the STO band structure, where
 the electrons of the LXO/STO interfaces mainly reside, we consider 
three parabolic bands schematizing the bottom of the $t_{2g}$-like bands of Ti.
The lowest ($d_{xy}$) band has light isotropic
mass $\sim 0.7 m_0$ ($m_0$ is the bare electron mass) while 
at an energy $\Delta\approx 50$ meV above there are two ($d_{xz}$ and $d_{yz}$) bands with the same light mass in one direction and 
a heavy mass as large as $20 m_0$ in the other 
\cite{mattheis,popovic,delugas,ARPES1,ARPES2,king1,salluzzo}. For simplicity, we 
neglect all mixings among the three bands \cite{notamixing}. 

The lowest band has therefore isotropic dispersion
\begin{equation}
\varepsilon_\pm(k)=\frac{\hbar^2k^2}{2m}\pm\alpha k
\label{hamaniso}
\end{equation}
which is the prototypical model for 2DEGs with Rashba coupling $\alpha$ (see, e.g., Ref.  \cite{libroSO}) and gives rise to a 2D isotropic band structure composed of 
two branches [solid black curves in Fig. \ref{fig-dos} (b),(c)] split by $\Delta_k=2\alpha\vert k\vert$, 
with a minimum $-\varepsilon_0\equiv -m\alpha^2/(2\hbar^2)$ occurring on a circle of radius $\alpha m$. 
The corresponding density of states (DOS) is
$N(\varepsilon)=N_0\equiv m/(\pi\hbar^2)$ for $\varepsilon >0$ and 
$N(\varepsilon)=N_0/\sqrt{1+\varepsilon/\varepsilon_0}$ for $-\varepsilon_0<\varepsilon \leq 0$.
The other two bands give rise to (strongly) anisotropic 
dispersions
\begin{equation}
\varepsilon_\pm(k)=\frac{\hbar^2k_x^2}{2m_x}+\frac{\hbar^2k_y^2}{2m_y}\pm 
\sqrt{\alpha_x^2 k_x^2+\alpha_y^2k_y^2}+\Delta,
\label{hamananiso}
\end{equation}
with $\nu\equiv m_y/m_x\sim 30$ and $\eta\equiv \alpha_y/\alpha_x$ for the $d_{xz}$ band and 
$\nu\equiv m_x/m_y\sim 30$ and $\eta\equiv \alpha_x/\alpha_y$  for the $d_{yz}$ band.
The DOS arising from Eq.~(\ref{hamananiso}) depends on $(\varepsilon-\Delta)/\varepsilon_0$ only and is
given in terms of complete elliptic integrals of the first and third kind (see Supplemental Material).
The precise form of $\alpha_x$ and $\alpha_y$, and therefore of $\eta$, should be determined
by first-principle calculations, which are beyond the scope of this Letter.
However, we argue that $\eta$ should be intermediate between two 
extreme cases: $\eta=1$ (isotropic Rashba coupling) and $\eta=1/\nu$  [{\it i.e.} $\alpha_{x,y}=1/m_{x,y}$ to 
reconstruct the relativistic form of the spin-orbit coupling: $({\bf v}\times {\bf \sigma} )\cdot \hat{\bf E}$].
The band bottom occurs at $\Delta-\nu \varepsilon_0$ in the former case and at $\Delta-\varepsilon_0$
in the latter (see Fig.~\ref{fig-dos}(b) and (c)).

In any case, $\alpha$ and $\alpha_{x,y}$ depend on the perpendicular electric field $E$ but, instead of the 
customary simple proportionality, we here take a non-linear expression $\alpha(E)=\tilde{\alpha}E/\left(1+\beta E \right)^3$ which is also 
valid at large fields and stems from 
a standard derivation of the Rashba coupling \cite{libroSO,schapers} (see also, {\it e.g.}, Eq. (2) in Ref. 
\cite{king2} and Supplemental Material of this Letter). Thus, $\varepsilon_0=\gamma E^2/\left(1+\beta E\right)^6$ [with 
$\gamma\equiv m\tilde\alpha^2/(2\hbar^2)$] is an increasing function of $E$ up 
to moderate-large values [$\sim 1/(2\beta) $]. The key point here
is that the electric field is directly related to the number of electrons in the plane
$E(n)=E(0)+n E'$ and therefore
the larger is $n$, the deeper is the energy minimum $\varepsilon_0\sim[E(0)+n E']^2$. 
This may render energetically convenient for the system to attract electrons to have 
a deeper energy minimum, where more electrons can be accommodated at lower energy. 
This downward shift of the band bottom may overcome the increase of the Fermi level
due to the increased $n$ and an overall decrease of $\mu$ may occur, leading to 
a negative compressibility. 

Filling the bands in Eqs. (\ref{hamaniso},\ref{hamananiso}) one easily gets
the general expression of $\mu$ as a function of $n=n_0+\delta n$ at a fixed 
gate potential $V_g$. At lower densities, when only the isotropic band is involved, one obtains
\begin{equation}
\mu(n,V_g)=
\begin{dcases}
\frac{n^2}{\left(2N_0\right)^2\varepsilon_0(n,V_g)}-\varepsilon_0(n,V_g)	&	\mu<0\\
\frac{n}{N_0}-2\varepsilon_0(n,V_g)	&	\mu\ge0.
\end{dcases}
\end{equation}
Experiments are usually performed at a fixed $V_g$ (and therefore at fixed $\delta n$)
and a stable system uniformly reconstructs the surface forming a uniform $n_0$ electron
density at the interface to avoid the polarity catastrophe. We now ask whether the system 
could be unstable and, at fixed $V_g$, could display a tendency to inhomogeneous
reconstructions and a negative compressibility upon varying $n$ (or, equivalently, $n_0$ 
since $\delta n$ is kept fixed).
 
The inverse compressibility $\kappa^{-1}=\partial\mu/\partial n$ at fixed $V_g$ for the isotropic band reads
\begin{equation}
\kappa^{-1}=
\begin{dcases}
\frac{n}{2N_0^2\varepsilon_0}-\frac{\partial \varepsilon_0}{\partial n}\left[\left(\frac{n}{2N_0\varepsilon_0}\right)^2+1\right]	&	\mu<0\\
\frac{1}{N_0}-2\frac{\partial\varepsilon_0}{\partial n}	&	\mu\ge0.
\end{dcases}
\end{equation}
and can be negative in both the low-density ($\mu<0$) and in the high-density ($\mu\ge0$) regime. The condition $\kappa<0$
naturally implies an electronic phase separation, {\it i.e.} a separation of the system into regions of different electronic densities
to be determined by the standard Maxwell construction on the $\mu$ {\it vs.} $n$ curve.

\begin{figure}
\includegraphics[angle=0,scale=0.3]{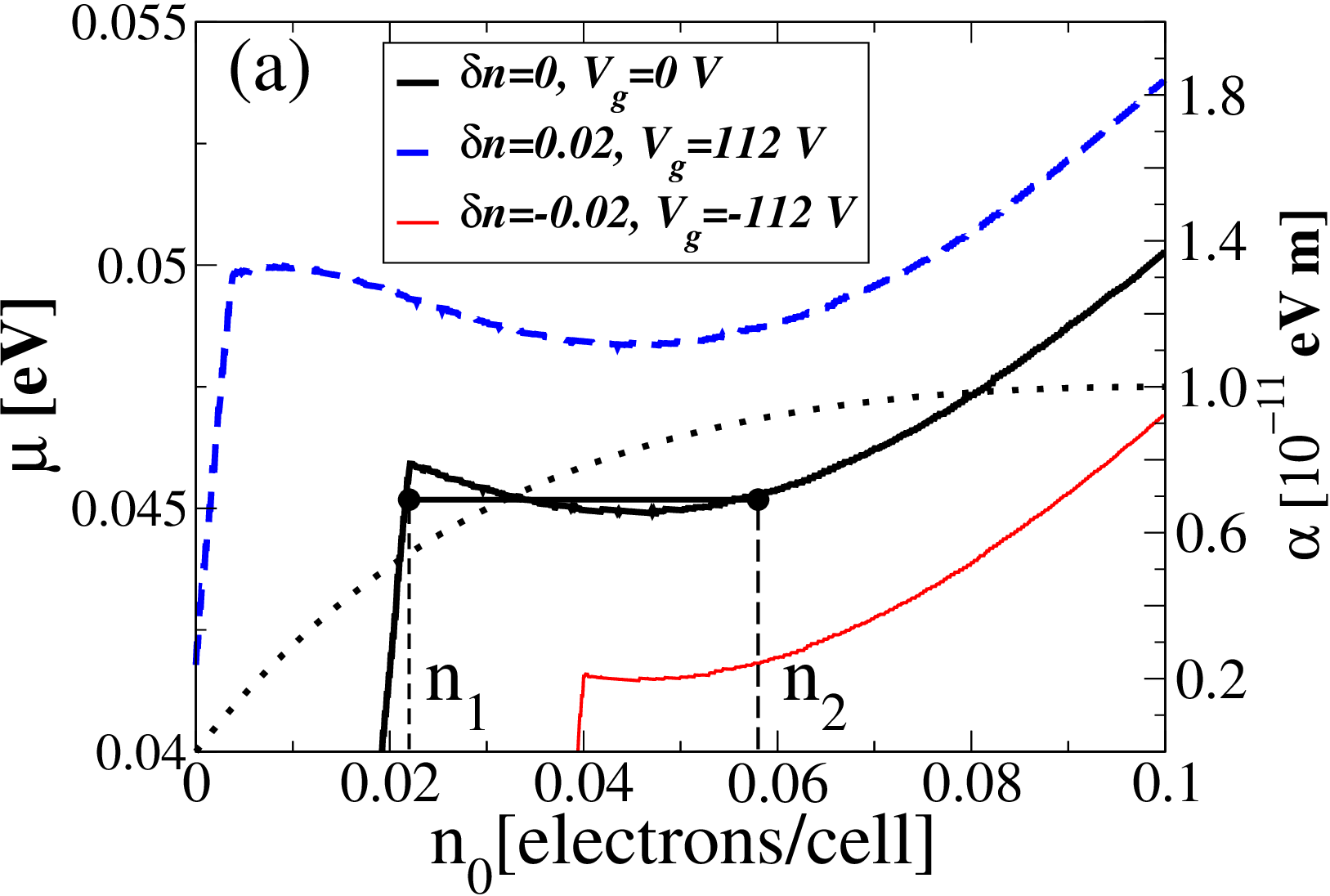}
\includegraphics[angle=0,scale=0.3]{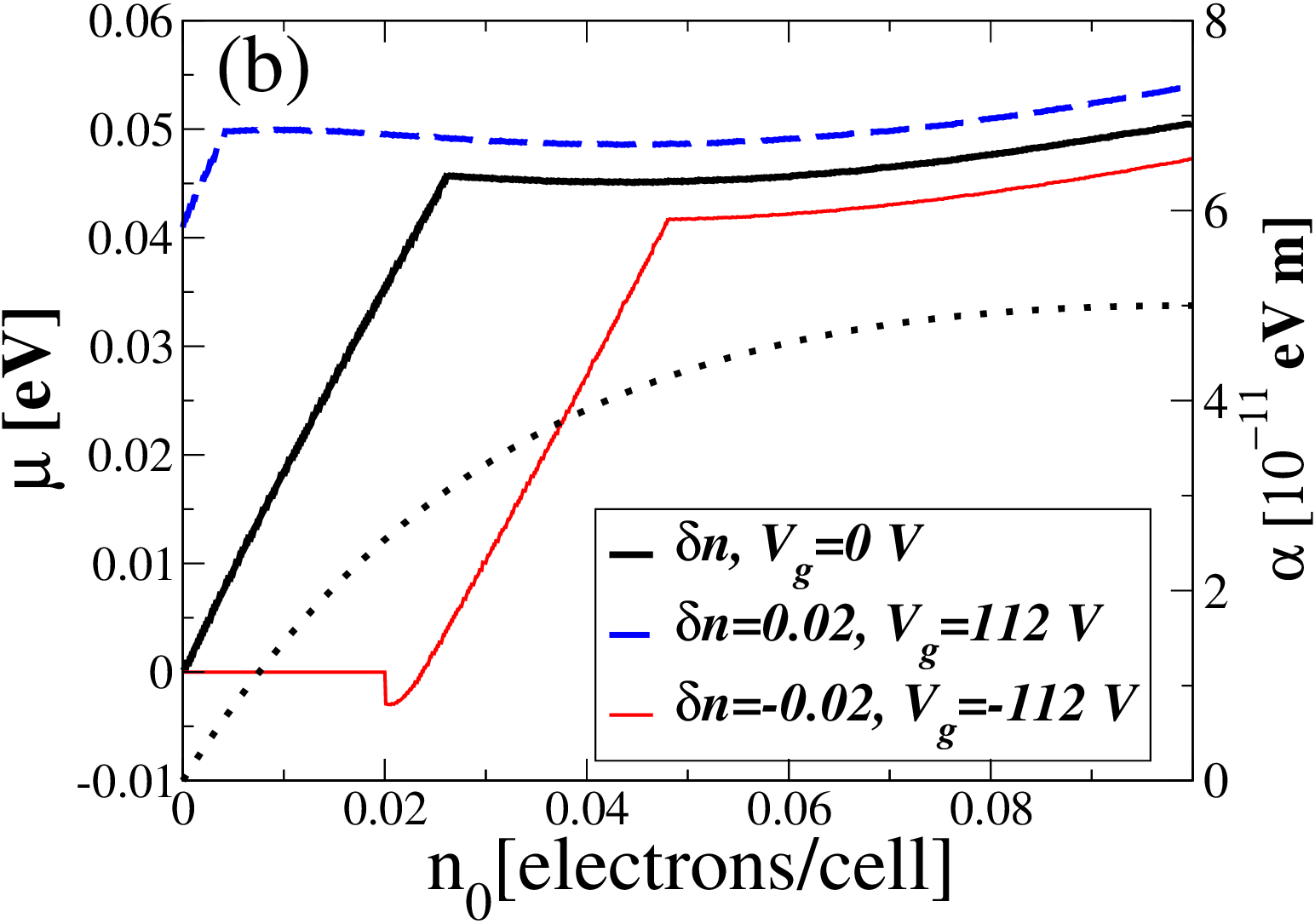}
\caption{(Color online) Chemical potential as a function of reconstructed electron density $n_0$ [$0.01$ electrons/cell correspond to $6,25\times 10^{12}$ cm$^{-2}$] for the three $t_{2g}$  bands of STO. The lowest band is isotropic ($\nu=\eta=1$) with mass $m\sim0.7m_0$. The two bands lying $\Delta=50$\,meV above are anisotropic with $\nu=30$ and $\eta=1$ (a) or $\eta=1/\nu$ (b). The legenda report the values of the additional electrons induced by gating. In (a) a Maxwell construction is also reported for the $V_g=0$ case. The dotted curves in (a) and (b)  (right axes) are 
$\alpha$ as a function of $n=n_0$ $(V_g=0)$.}
\label{figmuvsn}
\end{figure}

An analysis of Eq.~(4) for $\mu<0$ shows that if the electric field (and therefore $\alpha$)
stays finite when $n=n_0+\delta n \to 0$ ({\it i.e.} $E(0)\ne 0$), then $\kappa^{-1}\to-\partial\varepsilon_0/\partial n<0$ and
the system is always driven unstable at sufficiently low filling.
Although, depending on the value of $\alpha$, this latter 
instability could occur at very low densities [the $\delta n=-0.02$ curve (red online) in Fig.~\ref{figmuvsn}(b) displays a barely
visible negative initial slope around $n_0=0.02$ ({\it i.e. $n\approx0$})],
this observation might turn out to be relevant in 
physical systems like, e.g., MOSFETs or semiconducting heterostructures (see, e.g., Ref. 
\cite{ilaniold}) where $\alpha$ is small but $n$ also is very small.
In this regard, we considered the possibility that RSOC mechanism could be at
the origin of the negative compressibility recently observed in LAO/STO interfaces \cite{mannhart} at low filling.
We find that a negative compressibility  in the observed range $n\sim 10^{12}$ cm$^{-2}$ would require
values of $\alpha\sim 5\times 10^{-11}$\,eV\,m.

On the other hand, in the case $\mu\ge0$ of Eq.~(4) we can take the limit of not too large fields $\beta E\ll 1$ [and $E\approx nE'$]
and obtain $\kappa^{-1}\approx n^{-1}(n/N_0-4\varepsilon_0)$ which gives the simple condition $n/N_0<4\varepsilon_0$
for the negative compressibility to occur. For instance, at densities $n\sim10^{13}$ cm$^{-2}$,
$\kappa<0$ would require values as large as $\alpha\sim6\times10^{-11}$\,eV\,m.
It is also important to notice that a large DOS $N_0$ decreases
the positive contribution to $\kappa^{-1}$ and favors the occurrence of the instability.

Recent magneto-conductivity experiments in LAO/STO interfaces find substantial values for $\alpha\sim 10^{-12}\div10^{-11}$\,eV\,m
\cite{cavigliaprl,fete}. Similar values have been obtained for LTO/STO interfaces~\cite{espci3}. These values are about 5-8 times 
smaller than those estimated above to drive the system unstable. 
In the specific case of very small densities \cite{mannhart}, this quantitative discrepancy might well be due 
to the lack of many-body effects and disorder in our schematic model~\cite{notamannhart}).
The question then arises whether at higher densities $n\sim 2\div4 \times 10^{13}$ cm$^{-2}$,
with the anisotropic bands partially filled, LXO/STO interfaces displays
a negative compressibility at lower values of $\alpha$.
This question is addressed by Fig.~\ref{figmuvsn}, which is the main outcome of our analysis
and demonstrates the possibility of an inhomogenous electronic reconstruction
as soon as the anisotropic bands begin to fill. While 
the $\eta=1/\nu$ case [panel (b)] only becomes unstable for large values of 
$\alpha \sim 1\div5 \times 10^{-11}$\,eV\,m, the $\eta=1$ case [panel (a)]
displays electronic phase separation for quite realistic values of $\alpha\sim 0.1\div1 \times 10^{-11}$\,eV\,m,
well within the range of those found in Refs.~\onlinecite{cavigliaprl,fete} in LAO/STO.

To provide a clear understanding of this result, we consider the following expressions
\begin{gather}
n=2N_0\varepsilon_0\left[1+2f(\nu,\eta)\right]+\mu N_0+2(\mu-\Delta)N_0\sqrt{\nu},\label{mu3banda}\\
\mu=\frac{1}{1+2\sqrt{\nu}}\left\{\frac{n}{N_0}-2\varepsilon_0\left[1+2f(\nu,\eta)\right] +2\Delta\sqrt{\nu} \right\},
\label{mu3bandb}
\end{gather}
valid for $\mu>\Delta$ and $n>2N_0\varepsilon_0[1+2f(\nu,\eta)]+\Delta N_0$, respectively.
Here, $f(\nu,\eta)$ is a (rapidly) increasing function of the mass anisotropy $\nu$ (see Supplemental Material)
and gives a moderate [$f(30,1/30)\sim 3$] to strong [$f(30,1)\sim 90$] contribution
to the negative term in Eq.~(\ref{mu3bandb}). This shows that the RSOC-mediated instability
takes advantage from the mass anisotropy and a negative compressibility is now more likely to occur.

For $\mu<\Delta$ the linear growth of $\mu$ in Fig.~\ref{figmuvsn} shows that the DOS of the isotropic
$d_{xy}$ band is too small to promote an instability for the values of $\alpha (n)$ given by the dot-dashed curves.
On the other hand, as soon as the density $n_0+\delta n$ is large enough, the chemical potential 
eventually enters the anisotropic bands, the DOS rapidly increases (see, e.g., Ref. \cite{ilani}) and $\mu$ decreases 
with $n_0$ signaling a negative compressibility region.
For instance, the $V_g=0$ curve in Fig.~\ref{figmuvsn}(a)
has negative slope in the range $n_0\sim 0.02-0.04$ electrons/cell
($\sim3-5\times10^{13}$ cm$^{-2}$), which are typical values of as-grown systems.
Even for lower values of $n_0$, the instability eventually occurs upon increasing the density with $V_g$
(see, e.g., the $V_g=112$ V (blue online) dashed curve).
This leads to a thermodynamic phase separation
into regions of different densities (e.g., $n_1=0.022$ and $n_2=0.058$ in the $V_g=0$ curve of Fig.~\ref{figmuvsn}(a)) 
determined by a standard Maxwell 
construction \cite{phasesep}. For any given filling $n$, with  $n_1<n<n_2$,  the Maxwell
construction also determines the relative weight of the two phases at $n_1$ and $n_2$.

A few remarks are now in order. 
Firstly, we point out that the band structure of the 2DEG at the LXO/STO interfaces is substantially modified 
by the RSOC arising from the strong electric field due to the polarity catastrophe. Therefore,
it is not surprising that angle-resolved photoemission experiments \cite{ARPES1,ARPES2,king1}  
on STO/vacuum interfaces do not show the strong band splitting that one would expect from
the experiments of Refs.~\cite{cavigliaprl,fete}.

Secondly, we stress that additional mechanisms are also present in the real systems, which may cooperate with the 
intrinsic RSOC mechanism to produce inhomogeneous charge distributions.
Indeed, if the electronic system has a large (although still positive) compressibility,
impurities and defects more easily induce large inhomogeneities in such a ``soft'' and
largely fluctuating electron gas. In other words, they act as ``external fields'' on the density and 
may enhance the effects of the RSOC to locally induce phase separation. 
Previous work in strongly correlated systems \cite{CDG,GC,BTDG,GRCDK,BKCDG,CCCDGR}, also 
shows that electron-electron correlations and electron-phonon coupling favor
phase separation. In this regard, the recent analysis of RSOC within Hartree-Fock 
approximation \cite{agarwal}, can be a starting point for investigating the role of interactions on 
the RSOC-mediated instability.

Finally, we mention that our schematization of the LXO/STO interface in terms of a
2DEG should naturally be relaxed to account for the finite width of the interfacial electron gas
\cite{salluzzo,delugas,ARPES1,ARPES2,king1}. However, the introduction of subbands due to this finite 
width will surely increase the DOS and consequently decrease the $\mu$ {\it vs.} $n$ slope when the 
chemical potential progressively crosses the subbands. This will strongly favor the instability 
further reducing the RSOC required for it.

In conclusion, within our rather essential description of the LXO/STO interface, we 
find that electronic phase separation may likely occur at oxide interfaces, where 
substantial {\it density-dependent} electric fields can arise. 
We stress that a negative compressibility (cf. Ref.~\cite{mannhart}) {\it cannot} be explained by 
impurities, defects and so on, but is a distinct signature of an intrinsic mechanism
leading to an effective charge segregating attraction like the one proposed here.
We also show that (anisotropic) large masses (and subband splitting), isotropic RSOC, and low 
fillings favor the instability. This provides a natural framework accounting for the inhomogeneous 
phases observed in LXO/STO interfaces. It would be interesting to investigate the possible occurrence 
of electronic phase separation in cases where the RSOC is sizable and/or the density is very low, like 
in quantum wells, in boundaries of heavy metal alloys like Bi$_x$Pb$_{1-x}$, in 
reconstructed surfaces of Ag(111) \cite{ast,meier}, in MOSFET 
and semiconducting heterostructures at low densities, and in surface states
of topological insulators \cite{king2,yazdani}.

We thank  N. Bergeal, J. Biscaras, V. Brosco,
C. Castellani, C. Di Castro, T. Kopp, J. Lesueur, and R. Raimondi for  stimulating discussions and 
J. Mannhart for bringing Ref. \onlinecite{mannhart} to our attention.
We acknowledge financial support from ``University Research Project'' of the ``Sapienza''
University n. C26A115HTN.


\begin{thebibliography}{99}
\bibitem{notamixing} This mixing was recently claimed to play a role in transport \cite{ilani}, but it 
provides only minor quantitative corrections to our results. For simplicity we neglect it.
\bibitem{notamannhart} Several important differences make a precise comparison with the experimental data
of Ref. \onlinecite{mannhart} difficult. First of all the top-gating used in experiments act on the filling 
differently from the back-gating of our model in Fig. \ref{fig-dos}(a). Secondly the experimental inverse compressibility
values greatly depend on the samples, making a quantitative comparison hard to implement. 
Finally, some physical effects like electron-electron interaction and disorder become more important at low density,
thereby preventing a reliable quantitative use of our schematic model.
\bibitem{phasesep} The Maxwell construction only determines the densities of the coexisting phases, but says 
nothing on the size and structure of the phase separated regions. The separation in differently charged regions
may occur without paying a (large) Coulombic energy cost because the phase separated interface
is inserted between the conducting leeds of the gating potential and, more importantly,  because 
the polarity catastrophe and the related electronic reconstruction mechanism provide a suitable amount of countercharges,
which can be locally rearranged to reestablish a long-distance charge neutrality.
The possible formation of nano-, micro-,  or meso-scopic clusters
depends on several factors like the energy cost of cluster interfaces, screening length of the Coulomb electron-electron repulsion,
distance from the countercharges at the surface of the top layer. Examples of Coulomb-frustrated phase separation are 
considered in Refs.  \onlinecite{FPS}.
\end{thebibliography}
\end{document}



\author{S. Caprara}
\affiliation{Dipartimento di Fisica, Universit\`a di 
Roma ``La Sapienza'', P.$^{le}$ Aldo Moro 5, 00185 Roma, Italy}
\affiliation{ISC-CNR and Consorzio Nazionale Interuniversitario per le Scienze Fisiche della Materia, Unit\`a di Roma "Sapienza"}
 \author{F. Peronaci},
\affiliation{Dipartimento di Fisica, Universit\`a di 
Roma ``La Sapienza'', P.$^{le}$ Aldo Moro 5, 00185 Roma, Italy}
\author{M. Grilli}
\affiliation{Dipartimento di Fisica, Universit\`a di 
Roma ``La Sapienza'', P.$^{le}$ Aldo Moro 5, 00185 Roma, Italy}
\affiliation{ISC-CNR and Consorzio Nazionale Interuniversitario per le Scienze Fisiche della Materia, Unit\`a di Roma "Sapienza"}

\title{{\it Supplemental Material for}\\
Intrinsic instability of electronic interfaces with strong Rashba coupling}

\maketitle

\section{The Rashba Spin-Orbit Coupling}
A full {\it ab initio} derivation of the Rashba coupling is a quite challenging task, which goes beyond 
the scope of this Letter. Here, based on standard textbook arguments \cite{libroSO,schapers}, we simply remind 
that 
\begin{equation}
\alpha \propto \left<  \psi(z)\left| \frac{d}{dz} \left[  
\frac{1}{\varepsilon'(z)}-\frac{1}{\varepsilon'(z)+\Delta} \right] \right| \psi(z)\right>
\end{equation}
where $\varepsilon'(z)=\varepsilon +V(z)+E_{gap}$, $\varepsilon$ is the subband energy relative to the 
bulk conduction band, $V(z)$ is the band-bending potential, $E_{gap}$ is the band gap, 
and $\Delta$ is a measure of the spin-orbit coupling within a Kane ${\mathbf{k\cdot p}}$ approach 
\cite{kane}. Thus, since for a uniform electric field $E$ along $z$ the potential is $V(z)=-zE$, one 
obtains
\begin{eqnarray}
\alpha (E) &\propto& \left[\frac{1}{\left(\varepsilon +V(z)+E_{gap}\right)^2}-
\frac{1}{\left(\varepsilon 
+V(z)+E_{gap}+\Delta \right)^2}\right] \frac{dV(z)}{dz}
\approx \frac{ E\Delta}{\left(\varepsilon +E\overline{z}+E_{gap}\right)^3}\nonumber \\
&=& \frac{ \tilde\alpha E}{\left(1 +\beta  E \right)^3},   
\end{eqnarray}
where $\overline{z}\sim 2-6$ nm is the width of the region where the surface reconstruction occurs \cite{espci2}.
This expression has the standard linear behavior at small field, but saturates and then decreases at large fields. This latter
behavior is important to stabilize the system against unphysical unlimited growth of the electric field. Indeed, in the absence
of the denominator, the system would be unstable because it would be energetically (too) convenient to
attract large densities of electrons to increase enormously the electric field and the RSO coupling 
with the consequent formation of deeper and deeper
minima in the bands. This would eventually produce a negative compressibility. We emphasize, that the negative compressibility
found in the Letter is {\it not} of this type.

\section{The anisotropic Rashba band structure}
The real STO substrate has three bands, arising from the three $d_{xy}$, $d_{xz}$, and $d_{yz}$ orbitals.
Of course the atomic spin-orbit mixes these $t_{2g}$ levels (see, e.g., Ref. \cite{ilani} for 
an insight into possible consequences of this mixing), but here, we will neglect this mixing for the sake 
of simplicity and because it is unessential for the physical effects at issue in this Letter.
Then the 2DEG may be represented by the prototypical model Hamiltonian 
$H=\sum_{k,\sigma\varsigma}c_{k\sigma}^\dagger H_{\sigma\varsigma}(k)c_{k\varsigma}$, 
where $c_{k\sigma}^{(\dagger)}$ annihilates (creates) an electron with quasimomentum 
$k$ and spin projection $\sigma$, 
\beq
H_{\sigma\varsigma}(k)=\left(\frac{\hbar^2k_x^2}{2m_x}+ \frac{\hbar^2k_y^2}{2m_y}+ \Delta_{xy,xz,yz}\right)
\delta_{\sigma\varsigma}+\alpha_x k_x\sigma^y_{\sigma\varsigma}-\alpha_y  
k_y\sigma^x_{\sigma\varsigma},
\label{hamrashba}
\eeq
and $\sigma^{x,y}$ are the Pauli matrices. 
The lowest  band due to the $d_{xy}$ orbitals has 
a light isotropic mass, $m\sim 0.7 m_0$ ($m_0$ is the bare electronic mass) and $\Delta_{xy}=0$,
\begin{equation}
E^{iso}_\pm(k_x,k_y)=\frac{\hbar^2k_x^2}{2m}+\frac{\hbar^2k_y^2}{2m}\pm 
\sqrt{\alpha^2 k_x^2+\alpha^2k_y^2}
\label{hamiso}
\end{equation}
while  two anisotropic bands due to the $d_{xz}$ and $d_{yz}$ orbitals are present with a shift $\Delta_{xz,yz}\equiv\Delta$ at
 $k_x=k_y=0$. These latter
have a  heavy mass (as large as $20 m_0$) in one direction and a light mass in the other. 
These anisotropic bands 
have dispersions of the form
\begin{equation}
E^{aniso}_\pm(k_x,k_y)=\frac{\hbar^2k_x^2}{2m_x}+\frac{\hbar^2k_y^2}{2m_y}\pm 
\sqrt{\alpha_x^2 k_x^2+\alpha_y^2k_y^2}+\Delta,
\label{hamaniso}
\end{equation}
where $\Delta=50$ meV is the shift with respect to the $d_{xy}$ band and
$m_x=20 m_y=20 m$ (for the $d_{yz}$) or $m_y=20 m_x=20 m$ (for the $d_{xz}$).
The lowest band has four stationary points, which can be minima or saddle points 
depending on the values of $\nu\ge1$ and $\frac{1}{\nu}\leq\eta\leq1$. In the 
$\eta=1/\nu$ case the bottom of the band occurs at 
$-\varepsilon_0+\Delta$, where $\varepsilon_0\equiv m\alpha^2/(2\hbar^2)$, 
while for $\eta=1$ it is at $-\nu \varepsilon_0+\Delta$.

The DOS of the anisotropic $d_{xz,yz}$ bands
depends only on the ratio $\zeta=(\varepsilon-\Delta)/\varepsilon_0$ and its expression 
can be given analytically in terms of complete elliptic integrals of the first ($K$)
and third ($\Pi$) kind. Specifically, when $\eta^2\nu<1$
\begin{equation}
N(\zeta)=\theta[-\zeta (\zeta + 1)]{\cal{A}}_1(\zeta)+\theta[-(\zeta+\eta^2 \nu)(\zeta+1)]
{\cal{A}}_2(\zeta)
+\theta[\zeta] N_0 \sqrt\nu
\end{equation}
where 
\begin{eqnarray}
{\cal{A}}_1(\zeta)&=& 
\frac{2N_0\sqrt{\nu}}{\pi\sqrt{\eta^2\nu+\zeta}}
\left[-\zeta K\left(-\frac{\zeta (\eta^2 \nu-1)}{\eta^2 \nu+\zeta}\right) + 
(1 + \zeta) \Pi\left(\frac{\eta^2 \nu - 1}{\eta^2 \nu+ \zeta},\frac{-\zeta (\eta^2 \nu -1)}{\eta^2 \nu + 
\zeta}\right)\right] \\
{\cal{A}}_2(\zeta)&=& \frac{2N_0\sqrt{\nu}}{\pi\sqrt{- \zeta(\eta^2 \nu -1)}}
\left[-\zeta K\left(-\frac{\eta^2 \nu+\zeta}{\zeta (\eta^2 \nu - 1)}\right) \right.\nonumber \\
&&-\left.(1 + \zeta) \Pi\left(\frac{\eta^2 \nu+ \zeta}{\eta^2 \nu - 1}, \frac{\eta^2 \nu + \zeta}{-\zeta (\eta^2 \nu 
-1)}\right)\right] 
\end{eqnarray}
On the other hand, when $\eta^2\nu>1$
\begin{equation}
N(\zeta)=\theta[-\zeta (\zeta + \eta^2\nu)]{\cal{A}}_3(\zeta)+\theta[-(\zeta + \eta^2 \nu)(\zeta+1)]
{\cal{A}}_4(\zeta)
+\theta[\zeta] N_0 \sqrt\nu
\end{equation}
where 
\begin{eqnarray}
{\cal{A}}_3(\zeta)&=& 
\frac{2N_0\sqrt{\nu}}{\pi\eta\sqrt{1 + \zeta}}
\Pi\left(1-\frac{ 1}{\eta^2 \nu},\frac{-\zeta (1-\eta^2 \nu)}{\eta^2 \nu (1+ \zeta)}\right) \\
{\cal{A}}_4(\zeta)&=& \frac{2N_0\sqrt{\nu}}{\pi\sqrt{-\zeta(1-\eta^2 \nu)}}
\Pi\left(\frac{1+ \zeta}{\zeta}, \frac{\eta^2 \nu(1 + \zeta)}{\zeta (\eta^2 \nu -1)}\right)
\end{eqnarray}

\begin{figure}
\includegraphics[angle=0,scale=0.4]{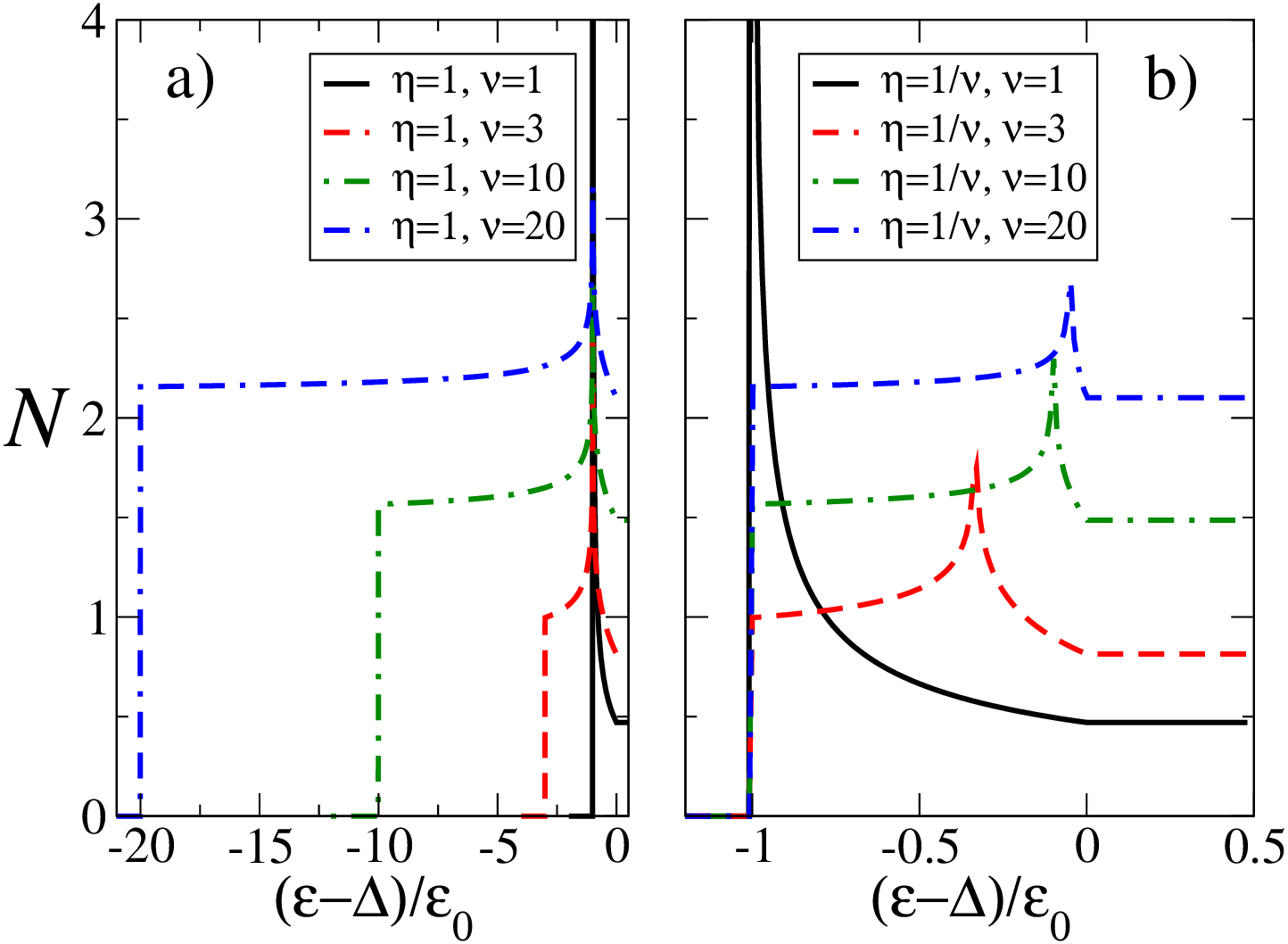}
\caption{(Color online) Density of states for the isotropic case (black, solid line) 
and for the anisotropic cases (the anisotropic bands do {\it not} include the
shift $\Delta$ to allow an easier comparison between the isotropic and anisotropic band
DOS): a)  Case with $\eta=1$ and $\nu=3$ (red online, dashed),   
$\nu=10$ (green online, dot-dashed),  and $\nu=20$ (blue online, dot-dot-dashed). b)  
Case $\eta=1/\nu$ and $\nu=3$ (red online, dashed), $\nu=10$ (green online, 
dot-dashed),  and $\nu=20$ (blue online, dot-dot-dashed).}
%
\label{fig-dos}
\end{figure}

Fig. \ref{fig-dos} reports the DOS for various values of $\nu$, in 
the extreme cases of $\eta=1$ (a) and $\eta=1/\nu$ (b). 

\begin{figure}
\vspace{2 truecm}
\includegraphics[angle=0,scale=0.45]{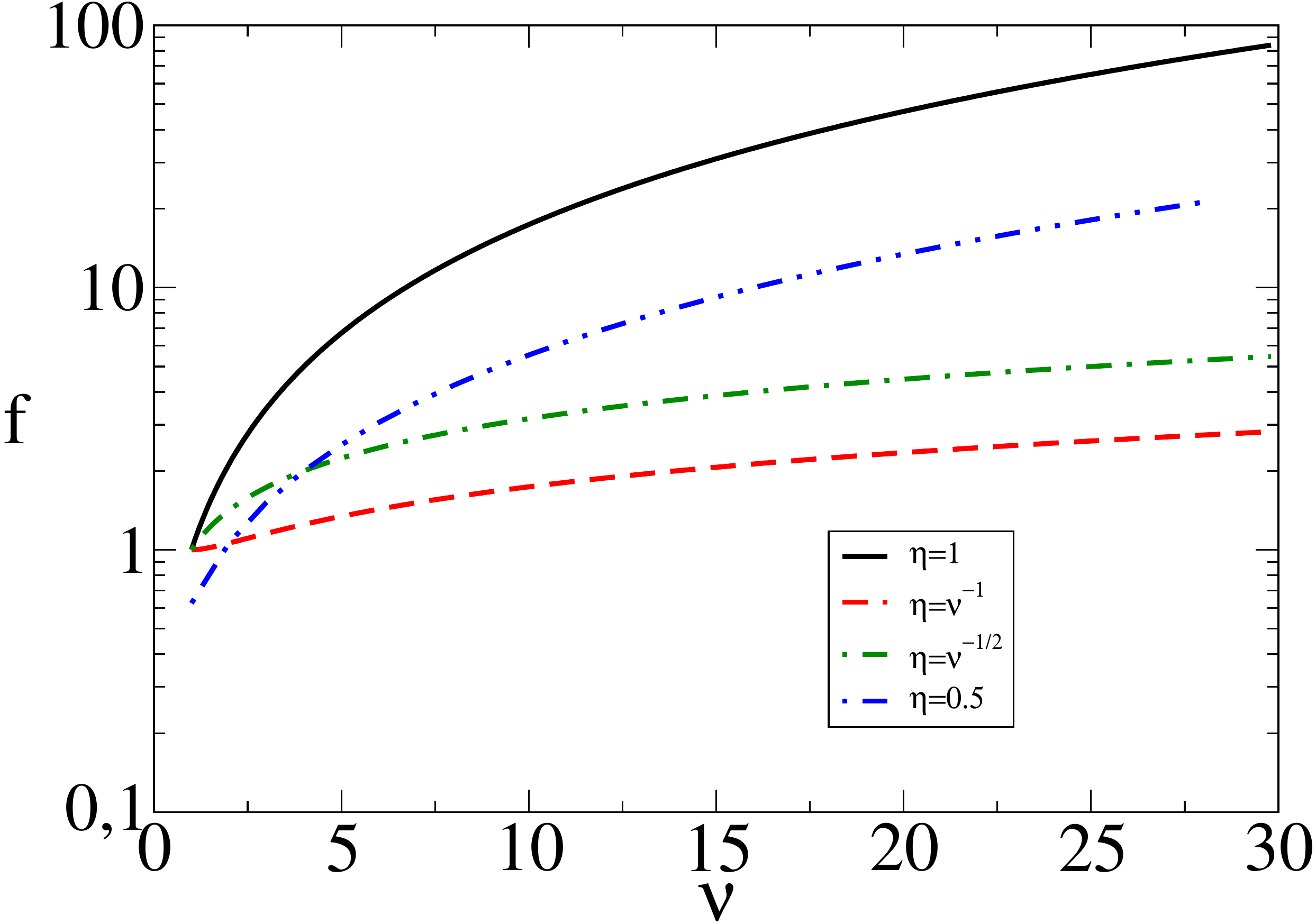}
\caption{(Color online) Enhancement factor $f(\mu,\nu)$ [cf. Eq. (\ref{muvsn1band})] 
for different values of the RSO anisotropy parameter $\eta$.}
\label{fnu}
\end{figure}

While the generic expressions for $\mu(n)$ are cumbersome,  for $\mu>\Delta$ the electron density inside 
each one of the anisotropic bands reads
$$n=(\mu-\Delta) N_0\sqrt{\nu}+2N_0\varepsilon_0f(\nu,\eta),$$
where 
$$2N_0 f(\nu,\eta)\equiv  \int^0_{\frac{\varepsilon_{min}-\Delta}{\varepsilon_0}}N(\zeta)d\zeta
$$
$\varepsilon_{min}(\nu,\eta)$ being the bottom of the lowest Rashba split anisotropic band. 
We insert the factor $2N_0$ to normalize this function to $ f (1, 1) = 1$ so that in the isotropic case 
with zero splitting $\Delta$ the Rashba term is 
$2N_0\varepsilon_0$, as previously found. $f(\nu,\eta)$ is a (rapidly) 
increasing function of the mass anisotropy $\nu$ (see Fig. \ref{fnu}). 
For the chemical potential one obtains
\begin{equation}
\mu=\Delta+\frac{1}{\sqrt{\nu}}\left[\frac{n}{N_0}-2\varepsilon_0 f(\nu,\eta)\right].
\label{muvsn1band}
\end{equation}
The negative term contains the function $f(\nu,\eta)\gg1$, which makes the condition 
$\kappa<0$ and the related electronic phase separation much easier to occur. As shown in Fig. \ref{fnu}, 
$f(\nu,\eta)$ depends on the anisotropy of the RSO term $\eta$. 
To reconstruct the relativistic form of the SO,
$({\bf v}\times {\bf \sigma} )\cdot \hat{\bf E}$, one 
should assume $\alpha_{x,y}\propto m_{x,y}^{-1}$, i.e., $\eta=1/\nu$. In this case, 
$f(\nu,1/\nu)\sim\sqrt{\nu}$ grows rather slowly with $\nu$. On the other hand, if 
the RSO term is isotropic ($\eta=1$) despite the mass anisotropy, 
$f(\nu,1)\sim \nu^{\frac{3}{2}}$ is a rapidly increasing function of $\nu$. The 
precise RSO coupling in real materials should be determined by first-principles 
calculations, which are beyond the scope of this Letter, but the behavior of 
$f(\nu,\eta)$ should be intermediate between these two extreme cases. Therefore, 
our analysis shows that the RSO-mediated instability can take a moderate or strong 
advantage from the mass anisotropy.

%
